\documentclass[fleqn,10pt]{wlscirep}
\usepackage[utf8]{inputenc}
\usepackage[T1]{fontenc}
\usepackage{graphicx}%
\usepackage{multirow}%
\usepackage{amsmath,amssymb,amsfonts}%
\usepackage{amsthm}%
\usepackage{mathrsfs}%
\usepackage[title]{appendix}%
\usepackage{xcolor}%
\usepackage{textcomp}%
\usepackage{manyfoot}%
\usepackage{booktabs}%
\usepackage{algorithm}%
\usepackage{algorithmicx}%
\usepackage{algpseudocode}%
\usepackage{listings}%
\usepackage{algorithm}
\usepackage{cellspace}

\floatname{algorithm}{Protocol}

\newcommand\cincludegraphics[2][]{\raisebox{-0.3\height}{\includegraphics[#1]{#2}}}

\newcommand{\ket}[1]{|#1\rangle}

\title{Q-COSMIC: Quantum Software Metrics Based on COSMIC (ISO/IEC19761) }

\author[1]{Francisco Vald\'es-Souto}
\author[2]{Hector G. Perez-Gonzalez}
\author[3,*]{Carlos A. Perez-Delgado}

\affil[1]{Universidad Nacional Aut\'onoma de M\'exico, Mexico City, M\'exico}

\affil[2]{Universidad Aut\'onoma de San Luis Potosí, San Luis Potosí, M\'exico}

\affil[3]{University of Kent, School of Computing, Canterbury, CT2 7NF, UK}

\affil[*]{c.perez@kent.ac.uk}

\begin{abstract}
Quantum engineering seeks to exploit quantum information  to build, among others, computing, cybersecurity, and metrology technologies. Quantum Software Engineering (QSE) focuses on the \emph{information processing} side of these technologies.
Historically, quantum (software) engineering has focused on development in controlled research environments and \emph{`in the small'}. As the field progresses, we should expect to see more large-scale quantum systems to be deployed as `real-world' products and services. An essential tool in (classical) software engineering and development has been \emph{software size} metrics. Calculating/estimating the size of a piece of software, to be developed or pre-existing, is an essential step in its engineering. Quantum software will be no different. Here we introduce Q-COSMIC, a technique for measuring the functional size of quantum software, based on the well-regarded COSMIC standard (ISO/IEC19761) for classical software.
\end{abstract}

\begin{document}

\flushbottom
\maketitle

\thispagestyle{empty}

\section*{Introduction}\label{sec1}

Recent years have seen a large increase in the rate of development of quantum information technologies. Products and devices are now being introduced into the market which aim to exploit quantum effects such as  superposition and entanglement for such diverse applications as metrology, power generation and storage, communication, and of course computation.

As with (classical) information technologies before them, their precipitated move from the scientific laboratory to industrial production has raised the need to develop proper engineering standards and techniques.

In an effort to stave off a repeat of the \emph{`software crisis'} of the 1960s, researchers have recently begun developing the field of \emph{quantum software engineering (QSE).} If software engineering is, as defined by the IEEE, \emph{``the application of a systematic, disciplined, quantifiable approach to the development, operation, and maintenance of software, as well as to the study of these approaches; that is, the application of engineering to software,''} then \emph{quantum} software engineering is the application of engineering to the development, operation, and maintenance of \emph{quantum} software.

Quantum software engineering is best understood as a super-set of (classical) software engineering, or as an extension of it, in order to accommodate software systems that have both a classical and a quantum component. Any  quantum information system of even modicum size is bound to have both intricate classical and quantum subsystems. QSE tools and techniques, therefore, need a way to analyse and portray both classical and quantum systems, and the interplay between both.

Moreover, a good quantum software engineering tool/technique/device should be identical/indistinguishable from a classical counterpart when the object of discourse is a purely classical software system. This is to ensure backward compatibility, ease-of-use, and keep a unified discourse language. 

That said, as has been observed before\cite{Perez-Delgado2020,Perez-Delgado2022}, quantum information is sufficiently distinct from classical information that is important to have language that distinguishes between the two even at a design level documentation.
In other words\cite{Perez-Delgado2022} \emph{``quantum software engineering should be as similar to classical software engineering as possible, but no more.''}

With the above in mind, we introduce here Q-COSMIC, a quantum extension of the  COSMIC (ISO/IEC 19761) standard for  a Functional Size Measurement Method (FSMM) for software. 

Software \emph{sizing} is an essential technique in software development, used in estimating and planning projects, measuring team productivity, evaluating
project performance, making benchmarks between technologies, suppliers, \emph{etc.} A (naive) measure of software size is simply its total lines of code. A more robust sizing method is based on the total functionality the software system provides its user-base---this is called a functional size measurement method (FSMM).

COSMIC (ISO/IEC 19761) is recognised as one of the leading standard for FSMM software sizing, due to its ease-of-use, large user base, and various advantages over previous generation methods\cite{Abran2010}.

Our extension of COSMIC aims to be \emph{minimal}. Q-COSMIC is indistinguishable from COSMIC when analysing purely classical projects. At the same time, Q-COSMIC has all the tools necessary to properly study, assess, and measure, quantum software projects. In this regard it achieves the goal set for quantum engineering mentioned above, as well as those argued in the \emph{Talavera Manifesto}\cite{Piattini2020}, \emph{e.g.}``QSE is agnostic regarding quantum programming languages and technologies'' and ``QSE embraces the
coexistence of classical and quantum computing.''

While this is not the first attempt to generalise COSMIC to work on quantum software\cite{Khattab2022}, this is the first time this done at the \emph{functional unit} level of abstraction, which we will argue later is the only logical and proper way to extend COSMIC.

\section*{Background}\label{sec2}

In this section we give a brief overview of the background topics needed to support our results. We begin with quantum computation and quantum information. 

\subsection*{Quantum Computation and Quantum Software }

Quantum computation aims to utilise quantum effects---such as superposition and entanglement---to manipulate information in a computationally advantageous way. The design and development of quantum algorithms, their composition into quantum software, and their use in solving computational problems that are infeasible classically, are topics that could fill several books; see \emph{e.g.} Nielsen and Chuang's excellent textbook\cite{nielsen2002quantum} in the area.

Here, we will limit ourselves to point out a few essential facts for our analysis. The first is the central concept of quantum information, whose fundamental unit is the quantum bit, or \emph{qubit}. A qubit, unlike its classical counterpart, can exist in a \emph{superposition} of its basis states $0$ and $1$. Quantum information can only be evolved (generally speaking) in a \emph{unitary} or \emph{reversible} manner.  In a fundamental physical sense, classical information and quantum information are \emph{incompatible} data types: classical information cannot be put in superposition, quantum information cannot be cloned, and cannot be evolved non-reversibly.

As with the programming languages concept of \emph{casting} it is possible to convert between quantum and classical data types. The process of creating a quantum state based on a classical one is known as \emph{state preparation}. While physically this process can be quite laborious, conceptually it is pretty straightforward: given a classical state $0$ ($1$) a quantum state $\ket{0}$ ($\ket{1}$) is produced.

The reverse procedure---creating a classical bit from a qubit---is called \emph{measurement} and is slightly more complicated. Given a quantum state $\ket{\phi} = \alpha\ket{0} + \beta\ket{1}$, a measurement will destroy the quantum bit, and create a classical bit whose state will be $0$ with probability $|\alpha|^2$ and $1$ with probability $|\beta|^2$. Hence, this procedure is generally lossy and irreversible. 

It is important to stress a deep difference between  the distinctions among classical data types, and the distinction between classical and quantum information types. The various classical data types (\emph{e.g.} \texttt{float}, \texttt{int}, \texttt{boolean}) are merely constructs of the software designer/programmer and exist solely for their convenience. The difference between classical and quantum bits is \emph{physically fundamental}. For instance, the laws of physics themselves prohibit the cloning of a quantum bit without first (explicitly or implicitly) \emph{casting} (transforming) it into a classical bit.

This distinction, and its importance for quantum software engineering, has been alluded to before, notably in the design of the quantum software design and modelling language Q-UML\cite{Perez-Delgado2020,Perez-Delgado2022}, which we discuss next.

\subsection*{Q-UML}

The Unified Modelling Language (UML) is a graphical language for designing, presenting, modelling, and analysing software at a high-level of abstraction. While it may be used to develop software under any paradigm, UML itself heavily uses the concepts and lexicon of object oriented development. In UML an \emph{object} (the encapsulation of a data-type and its associated operations) interacts with other objects via messages. Every object is a member of a \emph{class}. Classes are organised into hierarchies (posets) of \emph{inheritance}, a \emph{subclass} being a more specialised/specific class of a more general/abstract \emph{superclass}. As an example, a class \emph{car} can have a class \emph{motor vehicle} as its \emph{superclass}, making \emph{car} a \emph{sublcass} of the class \emph{motor vehicle}. In turn, the  black, 4.6 L V8 2005 Fort Mustang GT belonging to an author's family member is an instance, or \emph{object}, of class \emph{car}.

UML is a graphical language allowing for various types of diagrams. Of particular importance to us are class diagrams, which show the static relationship between the various classes in a piece of software; the sequence diagram, which shows how objects of the various classes interact with one another dynamically over the execution of said software; and use-case diagrams which show the software interacts with its intended users. For a more in-depth discussion of UML refer to one of many excellent textbooks on the area\cite{umlbook}.

Q-UML\cite{Perez-Delgado2020,Perez-Delgado2022} is an attempt to generalise UML to allow for the consideration of quantum software projects. Q-UML aims to be a \emph{minimal} extension that changes UML only as much as necessary to accomplish the above task. In particular Q-UML diagrams, and UML diagrams are indistinguishable when addressing purely classical software (in this sense, Q-UML is fully backwards compatible). 

The Q-UML extensions to UML are all meant to address the central issue of engineering quantum software (discussed earlier): quantum information and classical information are fundamentally different; and they are different to a level that goes beyond mere \emph{implementation details}. The use of quantum or classical information in a software module is a core design choice. Hence, quantum and classical data-types and procedures need to be explicitly delineated in a software design document.

The core principles of Q-UML---the need to differentiate quantum and software information types, and the insistence on modifying classical design tools only inasmuch as needed to do so---have been widely accepted by the community as sound, see \emph{e.g.} \emph{The Talavera Manifesto}\cite{Piattini2020}. 

Q-UML is important for us here for two reasons. The first is that in extending COSMIC to create Q-COSMIC we will also adhere to the principles delineated in the paragraph above. The second is that COSMIC is a measurement technique used on or applied to software design documents---most notably UML diagrams. It is therefore natural to showcase Q-COSMIC by applying it to Q-UML diagrams (as we shall do in the sections that follow).

Very briefly, Q-UML extends UML in two essential ways. First, it provides (graphical) language to distinguish between classical and quantum data-types, objects, classes, \emph{etc.} Classical structures are represented in regular typeface (or italics), and using single-lines (or single dotted-lines). In contrast, quantum structures are represented in bold typeface (or bold italics), and using double-lines (or double dotted-lines).

Q-UML also provides rules for how quantum and classical structures can interact with one another, and for how quantum/classical modules interact with UML concepts such as inheritance, aggregation, and communication. The two essential rules for our purposes here are 1) a structure  (object/class) that encapsulates quantum structures or quantum-data types will itself be considered a quantum structure and 2) a classical structure may only send and receive classical information messages, whereas a quantum structure may send and receive both quantum and classical information messages. 

Beyond providing a syntactic set of rules to describe quantum software, Q-UML sets forth a series of design rules and principles to be followed\cite{Perez-Delgado2020,Perez-Delgado2022}:
\begin{description}
\item[Quantum Classes (QCP):] Whenever a software module makes use of quantum information, either as part of its internal state/implementation, or as part of its interface, this must be clearly established in a design document.
\item[Quantum Elements (QEP):] Each module interface element (\emph{e.g.} public functions/methods, public variables) and  internal state variables can be either classical or quantum, and must be labelled accordingly (this applies to both variables, and functions).
\item[Quantum Supremacy (QSP):] A module that has \emph{at least} one quantum element is to be considered a quantum software module, otherwise it is a classical module. Quantum and classical modules  should be clearly labelled as such.
\item[Quantum Aggregation (QAP):] Any  module that is composed of one or more quantum modules will itself be considered a quantum module, and must be labelled as such.
\item[Quantum Communication (QCP):] Quantum and classical modules can communicate with each other as long as their interfaces are compatible, \emph{i.e.} the quantum module has classical inputs and/or outputs that can interface with the classical module.
\end{description}\label{pg:QUML_principles}

The above principles are important to us, as they are followed verbatim by Q-COSMIC. For a fuller discussion of Q-UML, including a reasoned motivation behind the above principles, please see the previous work on that modelling language\cite{Perez-Delgado2020,Perez-Delgado2022}. In the next section we will cover the essentials of software size metrics.

\section*{Metrics in Software Engineering }

Throughout software engineering development, software size metrics have been crucial in identifying challenges and formulating solutions through quantitative analysis.
However, many metrics put forth by practitioners and researchers primarily rely on intuition or empirical evidence, often resorting to simple entity counting and neglecting the mathematical properties and scales of numbers---an approach with various disadvantages\cite{Abran2010}. Additionally, as have been pointed out before\cite{Valdes-Souto2022}, except for functional size, most metrics tend to be descriptive or qualitative rather than quantitative. 

The measuring of software size in terms of its Functional User Requirements (FURs) has emerged as a way to provide more quantitative, standardised, and rigorous software size analysis
There are currently five competing standards based on this approach: COSMIC (ISO/IEC 19761), IFPUG (ISO/IEC 20926), MKII (ISO/IEC 20698), NESMA (ISO/IEC 24570), and FISMA (ISO/IEC 29881). For reasons outlined in the next section, we have chosen to base our work on the COSMIC standard.

As Quantum Software Engineering (QSE) continues to grow, it will be crucial to integrate the insights already gained from classical software engineering. There are clearly broad industrial application of quantum software projects. The importance of QSE in managing these projects effectively, is equally clear. Hence, establishing robust and scientifically valid engineering metrics for QSE from the outset is of great importance, both in academia and industry.

\subsection*{COSMIC Functional Size Measurement Method }

\begin{table}[]
    \centering
    \begin{tabular}{|Sc|b{(\textwidth - 4.0cm)}|}
    \hline
     Symbol & Interpretation \\
    \hline
    \cincludegraphics[width=1.8cm]{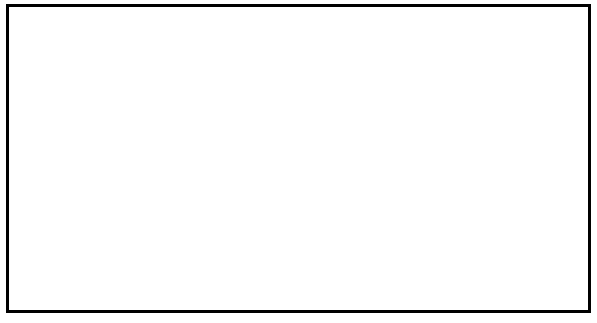} & The piece of classic software to be measured (box with thick outline) i.e. the definition of a measurement scope. \\
    \hline
      \cincludegraphics[width=1.8cm]{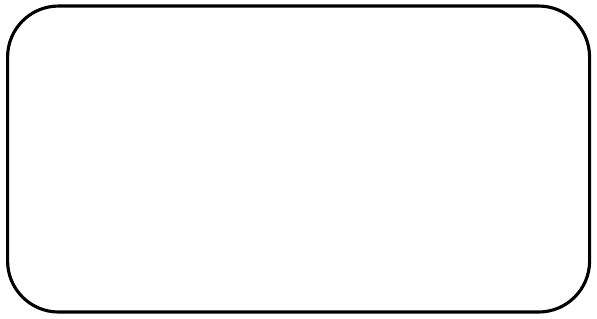}  & Any functional user of the classic software being measured. \\
    \hline
       \cincludegraphics[width=1.8cm]{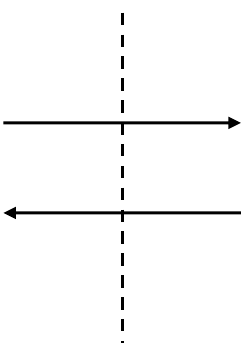}  & The arrows represent all the movements of data crossing a boundary (the dotted line) between a functional user and the classic software being measured. \\
    \hline
        \cincludegraphics[width=1.8cm]{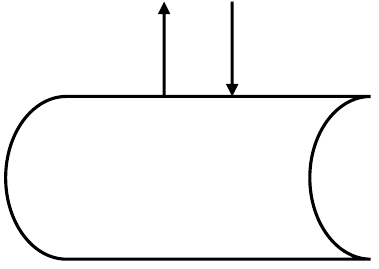} & The arrows represent all the movements of data between the classic software being measured and persistent storage.  \\
    \hline
    \end{tabular}
    \caption{\textbf{COSMIC Context Diagram Symbols:} This table shows the four basic symbols that appear on COSMIC diagrams, and their semantic interpretation.}
    \label{tab:cosmic-symbols}
\end{table}

The COSMIC Functional Size Measurement Method, as outlined in the COSMIC Measurement Manual \cite{Abran2021}, provides a systematic approach to gauging the functional size  of software through the use of Functional User Requirements (FURs). 

The COSMIC software measurement process is neatly divided into three phases: \emph{strategy}, \emph{mapping}, and \emph{measurement.}

The strategy phase, has  three main goals. The first is to determine the purpose of performing a  COSMIC measurement or analysis. 
The second is to clearly define who the functional users are. 
The third goal is to define the scope of the Functional User Requirements (FURs) to be measured. 
Succinctly, a FUR is a service or functionality that the software provides to the user, \emph{e.g.} a bank application may have \emph{`add a new account'}, and \emph{`search and return all accounts based on a certain search pattern'} among its FURs.

With the information obtained above, a context diagram is then  generated. Tab.\ \ref{tab:cosmic-symbols} summarises COSMIC's symbolic language for such diagrams.

The next step is the \emph{mapping} phase. During this stage, the FURs are mapped onto a COSMIC Generic Software Model (the general modelling approach used by COSMIC, as briefly described here, and in more detail in the COSMIC user manual\cite{Abran2010}). Afterwards, the data movements required to provide specific functionality for software are identified.

Finally, the measurement phase consists of counting the \emph{data movements}. COSMIC identifies four possible data movements, as detailed in Tab.\ \ref{tab:movements}. Each of these movements represents a functionality, and increases the size of the analysed software project/system. This allows COSMIC software size measurements to be independent of implementation details, such as programming language, total lines of code, \emph{etc.}

For more detailed information on the COSMIC standard and method, please refer to the COSMIC measurement manual\cite{Abran2021}, or the official COSMIC website at \href{www.cosmic-sizing.org}{www.cosmic-sizing.org}

The COSMIC Functional Size Measurement Method (FSMM) is not without criticisms or drawbacks. Notably, FSMM is based on software design (\emph{e.g.} UML) diagrams, which in some software development cases may be non-existent or come later in the development process than ideal\cite{Desharnais2003, Santillo2011}.
That said, COSMIC  is recognised as one of the leading standards for functional software sizing due to its ease-of-use, large user base, and various advantages over previous generation methods\cite{Abran2010}.

In classical software engineering, it is generally accepted that software sizing is an essential technique in software development, used in estimating and planning projects, measuring team productivity, and evaluating
project performance, making benchmarks between technologies, suppliers, \emph{etc.} 

Yet, quantum-software has so far been developed without many of these tools. One reason is that historically and to this moment, quantum software is designed and developed `in the small.' Quantum software, for the most part, is designed to be run on just \emph{one} particular hardware platform, solve \emph{one} particular problem, for \emph{one} particular user (-group). This is likely to change to in the near- to mid-term future.

A second reason is, plainly, the lack of quantum software engineering tools; with the introduction of Q-COSMIC, in the next section, we aim to help remedy this.

\section*{Q-COSMIC}

Q-COSMIC, introduced here, is meant to be an extension or super-set of the COSMIC (ISO/IEC19761) standard. 
For a purely classical piece of software, an analysis performed by either COSMIC or Q-COSMIC ought to be identical. Q-COSMIC, however, allows for the measurement of  quantum  software. In other words, a key goal is to keep Q-COSMIC maximally \emph{backward-compatible} with COSMIC. 

Briefly and intuitively we can understand the changes going from COSMIC to Q-COSMIC  as follows. First, the basic process and methodology is unchanged. However, Q-COSMIC does introduce extra syntactic elements, and semantic rules, as detailed in this section. The newly-added syntactic elements allow us to portray quantum data, functional units, and information flow. Q-COSMIC maps the data-movement rules of COSMIC onto quantum data movement (\emph{e.g.} just as reading/writing from/to external storage incurs a software size increment in COSMIC, so does  reading/writing from/to quantum storage in Q-COSMIC).

Less intuitively obvious is the fact that Q-COSMIC imposes a cost on data movement between \emph{classical} and \emph{quantum} functional units. In other words, whenever a quantum data-type is instantiated using classical information (state preparation), or a classical data-type is instantiated using quantum information (measurement), this incurs a software size increment in Q-COSMIC. Crucially, this analysis is done at the \emph{functional} level (as opposed to an implementation level---see the end of this section for a full discussion on this last point).

Q-COSMIC follows the same syntactic rules, as well as the basic axiom of minimally changing existing software engineering tools\footnote{The axiom reads textually \emph{``Quantum software engineering should be as similar to classical software engineering as possible, but no more.''}}, adopted by Q-UML\cite{Perez-Delgado2022}. Hence, the basic syntax of Q-COSMIC (above and beyond the COSMIC standard) can be summarised as follows:
\begin{itemize}
    \item Quantum information will be typeset in bold font when it is represented textually to set it apart from classical information. This applies to context diagrams.
    \item Quantum information will be distinguished in visual diagrams by the use of \emph{double lines}.
    \item When representing quantum objects, processes, etc., both bold type and \emph{double lines} are employed wherever possible and appropriate.
    \item In Q-COSMIC, every element is, by default, classical. A Data Group (DG) is designated as a Quantum Data Group (QDG) if and only if at least one of its components---such as an attribute---stores quantum information. 
    \item Data movements can be purely classical, purely quantum, or start as one and end as the other---this later referring to \emph{quantum state preparation}, and \emph{quantum measurement}.
    \end{itemize}

\begin{table}[]
    \centering
    \begin{tabular}{|Sc|b{(\textwidth - 4.0cm)}|}
    \hline
     Symbol & Interpretation \\
     \hline
      \cincludegraphics[width=1.8cm]{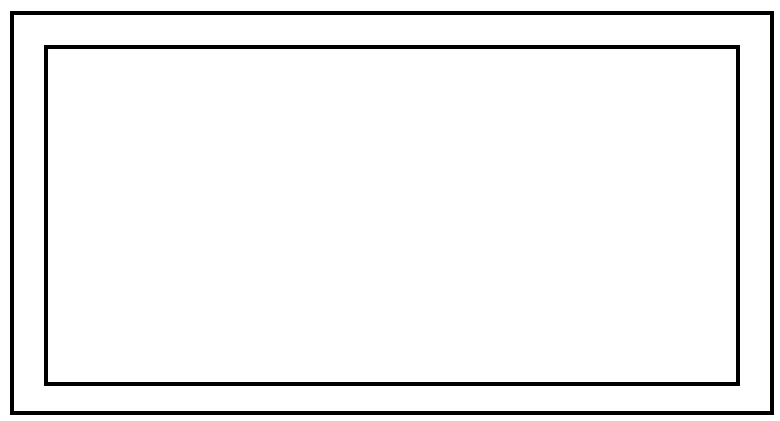} & The piece of quantum software to be measured (box with thick double line) i.e. the definition of a measurement scope \\
    \hline
       \cincludegraphics[width=1.8cm]{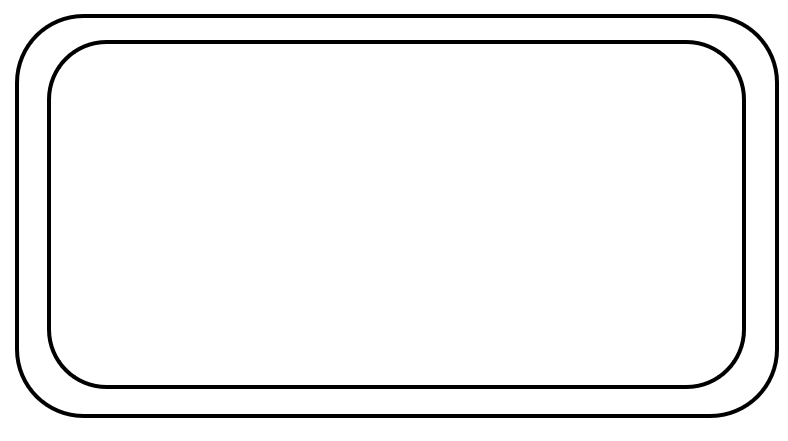} & Any functional user that is itself quantum in nature. \\
    \hline
     \cincludegraphics[width=1.8cm]{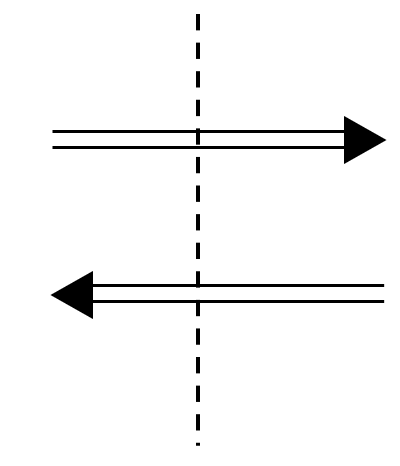} & The arrows represent all the movements of quantum data crossing a boundary (the dotted line) between a functional user and the quantum software being measured. \\
    \hline
      \cincludegraphics[width=1.8cm]{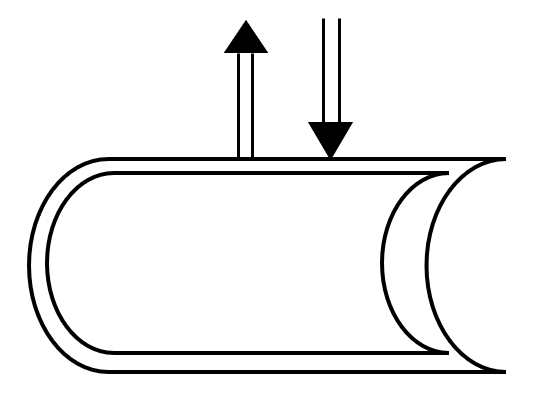} & The arrows represent all the movements of quantum data between the quantum software being measured and persistent storage.  \\
    \hline
    \end{tabular}
  \caption{\textbf{Q-COSMIC Context Diagram Symbols:} In addition to the symbols detailed in Tab.\ \ref{tab:cosmic-symbols}, Q-COSMIC introduces the four symbols presented here.}
    \label{tab:qcosmic-symbols}
\end{table}

Q-COSMIC introduces new syntactic elements to its pictorial language as detailed in Tab.\ \ref{tab:qcosmic-symbols}. In short, these new symbols allow us to portray quantum software, functional users, data, and storage, and to distinguish them from their classical counterparts.

\begin{figure}
    \centering
    \includegraphics[width=15cm,height=8cm]{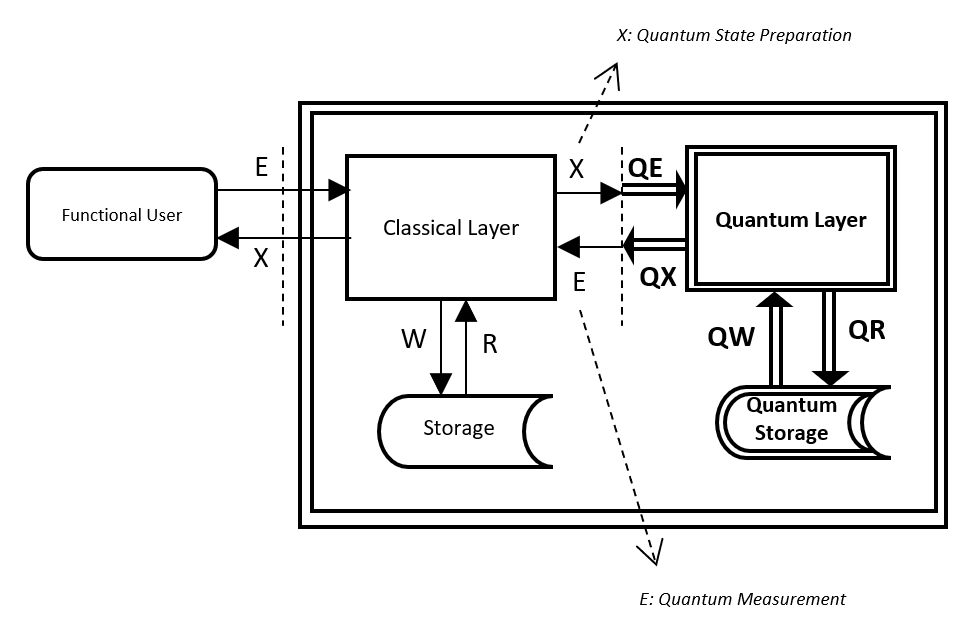}
    \caption{\textbf{Q-COSMIC Context Diagram:} This figure shows a generic quantum software context diagram in Q-COSMIC. It has two software layers: a classical, and quantum one. There is data movement between the user and the classical layer, and between the classical layer and the quantum layer. In this latter case, that data movement comes in the form of a quantum state preparation, and a quantum measurement. Finally, reading/writing to/from classical and quantum data storage is also portrayed.}
    \label{fig:generic}
\end{figure}

Additionally, Q-COSMIC introduces four new data-movement types. These are presented in Tab.\ \ref{tab:movements}. Each newly introduced data movement mirrors a classical analogue, and is meant to represent \emph{quantum data} movement. The COSMIC rules for data-movements, how they are counted towards software size (and when they are not) are extended \emph{in whole} to quantum data movements in Q-COSMIC.

\begin{table}[]
    \centering
    \begin{tabular}{|b{0.45\textwidth}|b{0.45\textwidth}|}
    \hline
    Data Movement (DM) Type in COSMIC & Quantum Data Movement (QDM) Type in Q-COSMIC\\
    \hline
    Entry (E): data movement that moves a classic data group from a functional user across the boundary into the functional process where it is required.  & \textbf{QEntry (QE)}: data movement that moves a quantum data group from a functional user across the boundary into the functional process where it is required.\\
    \hline
    eXit (X): data movement that moves a classic data group from a functional process across the boundary to the functional user that requires it.  & \textbf{QeXit (QX)}: data movement that moves a quantum data group from a functional process across the boundary to the functional user that requires it.\\
    \hline
   Read (R): data movement that moves a classic data group from persistent storage within reach of the functional process which requires it. & \textbf{QRead (QR)}: data movement that moves a quantum data group from quantum persistent storage within reach of the functional process which requires it.\\
    \hline
    Write (W): data movement that moves a classic data group from a functional process to persistent storage.  & \textbf{QWrite (QW)}: data movement that moves a quantum data group from a functional process to quantum persistent storage.\\
    \hline
    \end{tabular}
    \caption{\textbf{Data Movements:} The left hand column presents the four data movements pre-existing in COSMIC. The right hand column displays the four new data movements introduced in Q-COSMIC. Recall that each data movement incurs a cost, and hence increases the ultimate software size.}
    \label{tab:movements}
\end{table}

Q-COSMIC adheres to the same design principles as Q-UML, as described on Pg.\ \pageref{pg:QUML_principles}.

 Q-COSMIC adds the following \emph{rules} to the existing COSMIC standard. First is the use of the prefix \textbf{Q} to refer to quantum elements in writing. The second rule, alluded to above, is the introduction of the Q-COSMIC Function Point (QCFP), as the unit representing a single generalised (quantum or classical)  data movement . As an example, a write (W) to classical data storage is counted towards the size of a classical software system in COSMIC. Similarly, in Q-COSMIC a movement of quantum information from a quantum functional process to quantum data storage (QW), or to/from a quantum functional process  (QE/QX), are  considered data-movements, and incur a software size increments in Q-COSMIC.

In addition, Q-COSMIC makes the following changes to existing COSMIC definitions as they appear in the COSMIC User Manual V5, Part 1\cite{Abran2021}:

\begin{description}
\item [Data movement:] Base Functional Component which moves a single data group. The data group can be either classical or quantum.
\item[Functional process:] Elementary component of a set of Functional User Requirements, comprising a unique, cohesive and independently executable set of data movements. These can be either classical or quantum. 
\item[Software system:] A software system can be purely classical; or, it can be a quantum software system.
\item[Layer:] Partition resulting from the functional division of a software system. A quantum software system  will consist of at least one classical, and one quantum software layer.
\end{description}

The fact that both quantum measurements and quantum state preparation (that is, transferring information from a quantum data-type/module to a classical one) incur a software size increment in Q-COSMIC would seem to imply that every \emph{quantum gate} would incur a separate size increase cost in Q-COSMIC.

This is because quantum control, and hence the implementation of quantum gate, necessitates the interplay and cross-talk of both classical and quantum data types. This is particularly true when considering fault-tolerant quantum gates\cite{qft}.

However, Q-COSMIC as a super-set of COSMIC, 
measures data movements at \emph{functional} level of abstraction. The COSMIC manual lays out clearly\cite{CommonSoftwareMeasurementInternationalConsortiumCOSMIC2017}:  \emph{``The COSMIC Generic Software Model, as its name suggests, is a logical `model' that exposes units in which software processes data that are suitable for functional size measurement. The model does not intend to describe the physical sequence of the steps by which software executes nor any technical implementation of the software.''}

The  manual goes on to say that any data  manipulation that is  \emph{``internal to the functional process and that can be changed only by a programmer, or computation of intermediate results in a calculation, or of data stored by a functional process resulting only from the implementation, rather than from the FUR''} shall likewise not be counted in the functional size. 

So, as an example, a classical module messaging an integer to be factored to a quantum module would constitute a data transfer in Q-COSMIC; while the error syndrome being measured during the implementation of a fault-tolerant gate would not.

Before moving on, it is worth explaining the motivation behind some of the less intuitive or obvious design choices behind Q-COSMIC. For instance, quantum data read/writes from/to external quantum storage incurring a size cost in Q-COSMIC---just like classical read/writes incur a size cost in COSMIC---should seem like a natural choice. The further imposition, in Q-COSMIC, that quantum and classical functionality be separated into separate layers, and that data movement between these classical and quantum functional units incurs a software size increment cost, may seem arbitrary. The argument could be made that the classical pre- and post-processing of data before/after a quantum algorithm is part of the same functional unit. 

Furthermore, while it is unarguable that at an \emph{implementation} level quantum and classical software are indeed different, the argument has to be made that they are different at the \emph{functional} level, which is the abstraction level at which COSMIC and Q-COSMIC operate. However, this precise argument \emph{has been} made before, repeatedly\cite{Perez-Delgado2020,Perez-Delgado2022,Sodhi2018,Zhao2020}. Briefly, the argument goes as follows. Even at the software design abstraction level, quantum and classical modules or functional units are distinct and must be properly identified. This is because quantum and classical software modules obey fundamentally different logical rules; need to be developed using different techniques/languages/methodologies; and must be tested (separately) using different tools. This is not an external imposition, or arbitrary rule, this is so because of the fundamental laws of nature.

Given the above one might be tempted to argue that the development of classical and quantum functional units require different amount of efforts, and hence should be accounted for differently. We chose not to take this path in Q-COSMIC, for two fundamental reasons. The first, is for simplicity and maximum backwards compatibility with COSMIC. COSMIC, as a metric of software size, ultimately reports a single number. Q-COSMIC should do the same.

The second is more fundamental. Functional size, as reported by COSMIC, is \emph{not} a measure of algorithmic complexity, or the \emph{difficulty} of implementing a feature. COSMIC---and by extension Q-COSMIC---do not attempt to quantify the amount of human effort required to create a software product. What they do, is measure the \emph{functional} size of a software product.

For instance, COSMIC would not differentiate between a functional unit that sorts the entries on a  database of modicum size by the value of some field, and a functional unit that predicts the weather accurately months in advance. From COSMIC's perspective, they both provide a single functionality to the user. This is independent of the difficulty of implementing each functionality.

While implementing \emph{quantum} functionality may be even harder than the most challenging classical functionality, we argue that the difference in difficulty is one of merely degrees. All of this is not to say that Q-COSMIC can not report separately the functional sizes of  the \emph{classical} and \emph{quantum} modules of a software system if required.

Finally, it is worth addressing our choice to define Q-COSMIC to operate at the abstraction level of functional unit, rather than, say, quantum gates. We acknowledge that other authors have attempted to extend COSMIC to quantum software in this alternate way\cite{Khattab2022}. However, we would argue that the \emph{only} logical and proper way to extend COSMIC is to keep the accounting of quantum and classical software as homogeneous as possible. In other words, if classical software is measured at the functional level, then so should quantum software.

\begin{figure}
    \centering
    \includegraphics[width=8cm,height=7cm]{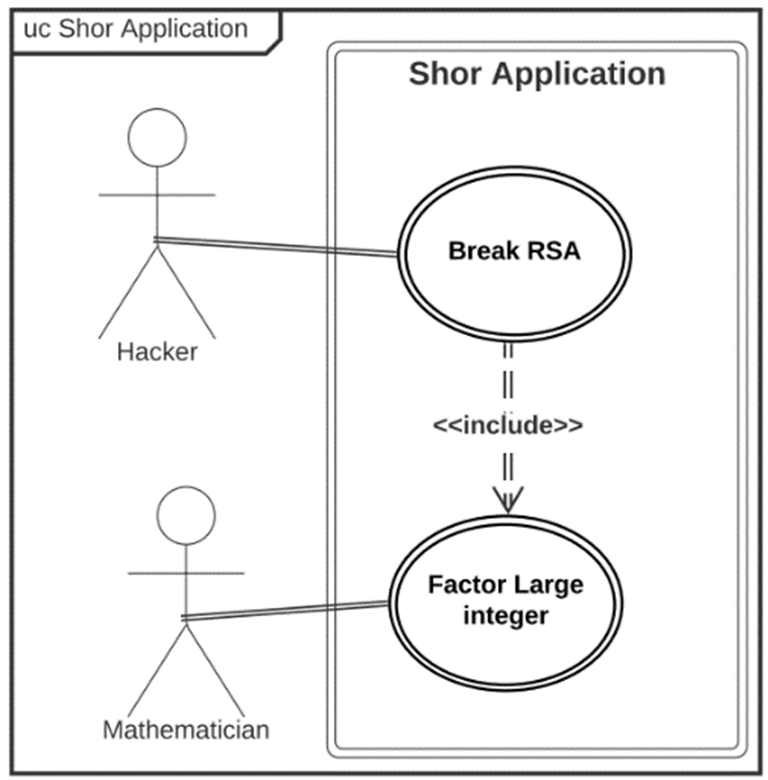}
    \caption{\textbf{Quantum software FUR’s:} This diagrams shows the Functional User Requirements (FURs) of a quantum large integer factoring software application. This figure is based on the ones used to exemplify the use of Q-UML\cite{Perez-Delgado2020,Perez-Delgado2022}.}
    \label{fig:F2}
\end{figure}

In order to give a better understanding of Q-COSMIC's syntactic and semantic rules, as well as principles, in the next section we will provide a Q-COSMIC analysis of a quantum software system designed to factor large integers.

\section*{Example: Measuring a Factoring Application with Q-COSMIC}

In this section, we will analyse an example quantum software system that factors large numbers using Shor's algorithm. Thhttps://www.overleaf.com/project/64a42106ef033ebd3015b963/detacheris software system has two potential users: mathematicians, interested in factoring large integers (for the sake of mathematical curiosity), and hackers who wish to break public-key cryptosystems, see Fig.\ \ref{fig:F2}. 

To do the above, we shall proceed through the three phases of a Q-COSMIC analysis in order.

\subsubsection*{The strategy phase}
As mentioned earlier, the three main goals of the strategy phase are to determine the purpose the  Q-COSMIC measurement, the scope of the functional requirements, and to identify who the functional users are.

\begin{description}
\item[Purpose:] this case, the purpose is to determine the  Functional Size Measurement (FSM) of the quantum software portrayed in Fig.\ \ref{fig:F2} which implements two use cases (the previously defined $UC_1$ and $UC_2$).
\item[Scope:] The scope of this measurement are the FURs related to $UC_1$ and $UC_2$.
\item[Functional Users:] The functional users are ``Matemathician'', and ``Hacker''. 
\end{description}

\begin{figure}
    \centering
    \includegraphics[width=15cm,height=8cm]{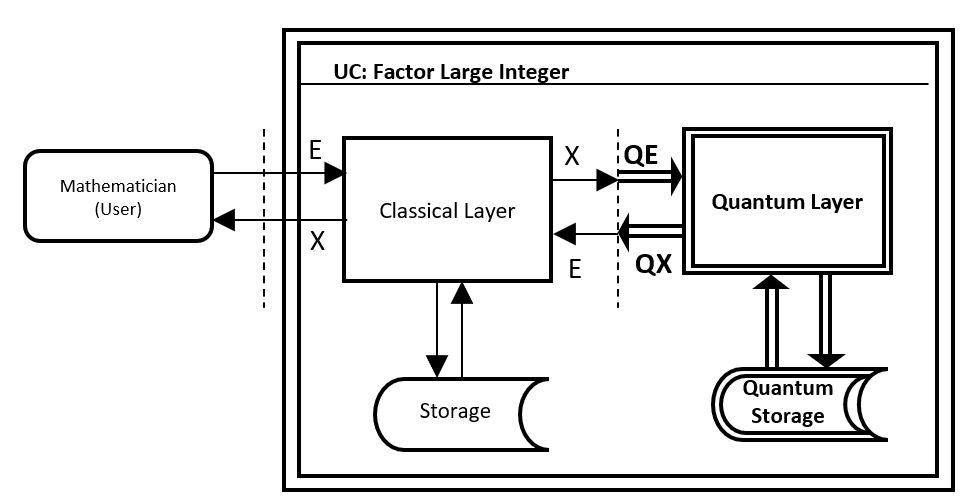}
    \caption{\textbf{Q-COSMIC Context Diagram for \emph{Factor Large Integer:}} This shows the context diagram for the first use case, $UC_1$ of the integer factoring software application. This diagram depicts six total data movements: two eXits (2X), two Entries (2E), one Quantum eXit (1QX), and one Quantum Entry (1QE).}
    \label{fig:UC1}
\end{figure}

With the above information we can generate a context diagram, which is shown in Fig.\ \ref{fig:UC1}. Note that since $UC_1$ corresponds to a quantum software system, Q-COSMIC rules state that two separate layers must be defined, a classical and a quantum one.

In order to provide its functionality, $UC_2$ is a functional user for the $UC_1$. This gives us the context diagram in Fig.\ \ref{fig:UC2}.

\subsubsection*{The mapping phase}
Recall, the purpose of this phase is to map FURs to the Q-COSMIC generic software model.

\begin{figure}[t]
    \centering
    \includegraphics[width=17cm,height=8cm]{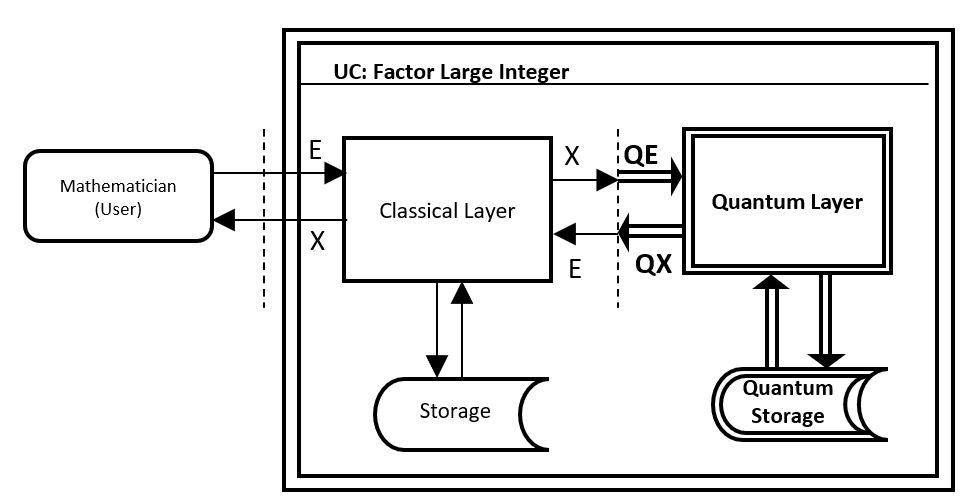}
    \caption{\textbf{Diagram Context for Break RSA Use Case:} Here we show the context diagram for the second use case $UC_2$ of our example software. The previous use case, $UC_1$ provides some of the functionality required by $UC_2$, and hence appears on this diagram. Importantly, while measuring the functional size of $UC_2$, we only count the  data movements newly introduced in $UC_2$, and \emph{not} the data movements pre-existing in $UC_2$. In this case, these consist of two Entries (2E), and two eXits (2X).}
    \label{fig:UC2}
\end{figure}

We established previously that each Functional Process (FP)
corresponds to a a UC. The next step hence is to identify the data movements in each of the use cases, $UC_1$ and $UC_2$. These are presented in Fig.\ \ref{fig:UC1}, and Fig.\ \ref{fig:UC2} respectively.

Note that during the measurement of $UC_2$ we do \emph{not} re-measure the size of $UC_1$, since it has already been accounted for. Since $UC_2$ is otherwise purely classical, its size in Q-COSMIC corresponds to the size that would be calculated using COSMIC.

\subsubsection*{The measuring phase}

Finally, after the Data Movements (DMs) in all use cases have been identified, in previous phases, the final phase is to simply count them up. Doing this calculation gives the Q-COSMIC functional size of $UC_1$ as:  $6$ QCFP, where QCFP is short for Q-COSMIC Functional Point. Whereas the Q-COSMIC functional size of $UC_2$ is $4$ QCFP $= 4$ CFPv5, where CFPv5 refers to the functional size units used in COSMIC version 5 (since this is a purely classical sub-system, its size in Q-COSMIC and in COSMIC coincides, as would be expected).

While the total functional size of the quantum software use cases are the ultimate numbers being sought, it will sometimes be useful to do a more granular analysis, separating the quantum and classical layers and measuring them separately. For $UC_2$, this granular analysis is irrelevant, since its own functionality is purely classical. In the case of $UC_1$, the quantum layer is of size $2$ QCFP. The size of the classical layer is $4$ QCFP $= 4$ CFPv5.

In summary, for the analysed FURs ($CU_1$ and $CU_2$), there are $6$ total functional points in the classical layer constituting $60\%$ of the total, and $4$ functional points in the quantum layer making up the remaining $40\%$. This granular analysis allows for the separation of estimation, productivity, and cost analysis for each type of software.

\section*{Conclusions}\label{sec4}

In this paper we have introduced Q-COSMIC, a proposed standard metric and tool for measuring and estimating the size of quantum software systems. Q-COSMIC extends the COSMIC (ISO/IEC 19761) standard using the same quantum software engineering principles developed earlier, namely for Q-UML\cite{Perez-Delgado2020,Perez-Delgado2022}.

Q-COSMIC aims to be a \emph{minimal} extension to COSMIC---retaining maximum backwards-compatibility, while still having the sufficient tools to analyse quantum software.
Q-COSMIC is grounded on the well-established principles that COSMIC is grounded on, while expanding on COSMIC sufficiently to consider  previously non-analysable software. Crucially, all results, tools, and advances made in the field of COSMIC measurement are applicable in Q-COSMIC with little or no modification.

Further, we demonstrate the principles and rules of Q-COSMIC, as well as its expressiveness and ease-of-use, by applying Q-COSMIC to an example quantum software system.

Q-COSMIC represents a novel contribution to the area of software metrics.
Moreover, the introduction of Q-COSMIC contributes to the broader goal of advancing QSE by establishing robust, scientifically valid metrics from the outset. This aligns with the principles outlined in the \emph{Talavera Manifesto}\cite{Piattini2020}, emphasising the importance of effective project management and industry-wide implementation of quantum software projects. 

The ultimate goal of  Q-COSMIC, like other QSE tools, is to help expedite the transition of quantum technologies from the laboratory, to a robust and growing industry.

\bibliography{sample}

\section*{Acknowledgements}

C.P-D.  would like to acknowledge funding through the  Quantum Communications Hub (EP/T001011/1), and from the Casper Association Academic Grants Program.

\end{document}